\documentclass[preprint, 12pt]{elsarticle}
\usepackage[version=3]{mhchem} 

\usepackage{xcolor}
\usepackage{gensymb}
\usepackage{tablefootnote}
\usepackage[normalem]{ulem}
\usepackage[bb=stix]{mathalpha}
\usepackage{url}
\usepackage[numbers]{natbib}
\usepackage{siunitx}
\begin{document}

\begin{frontmatter}

\title{Phase-Field Model of Solution and Stoichiometric Phases with Molar Volume Difference}
\author[1]{Chengyin Wu}
\author[1]{Yanzhou Ji}
\ead{ji.730@osu.edu}
\affiliation[1]{organization={Department of Materials Science and Engineering, The Ohio State                University},
            addressline={140 W 19th Ave}, 
            city={Columbus}, 
            state={OH},
            postcode={43210},
            country={USA}}

\begin{abstract}
Phase-field models have proven indispensable for deciphering the microstructure complexities inherent in multicomponent systems. The confluence of varying phase molar volumes presents unique challenges. Understanding the impact of molar volume differences on multiphase systems is of crucial significance, as it directly influences the system's thermodynamic and kinetic behavior, as well as resulting phase morphologies and distributions. In this study, we developed a phase-field model of solution and stoichiometric phases that can account for the molar volume differences. With the phase molar volumes taken from existing CALPHAD thermodynamic databases, we quantitatively investigated how the different molar volume settings would influence the growth rate and morphologies of stoichiometric $\theta^\prime$-\ce{Al2Cu} and $\beta$-\ce{Al140Mg89} precipitates in Al-based alloys. We anticipate that this approach can be applied to various materials systems with phase- and composition-dependent molar volumes for more accurate phase-field predictions.

\end{abstract}


\end{frontmatter}

\section{Introduction}
Stoichiometric compounds are present in the majority of material applications, ranging from high-refractory materials (\ce{Al2O3}, \ce{ZrO2}) to semiconductors (\ce{GaAs}). Many of these compounds can form from solution phases during materials synthesis, processing and operation, leading to changes in crystal structures, compositions, and notably, molar volumes. Understanding the impact of molar volume differences in multiphase systems is of crucial significance as it directly influences the system's thermodynamic and kinetic behavior. The molar volume ($V_{m}$), plays a pivotal role in governing phase transitions\cite{PhysRevA.45.7424, LIU2014540, PhysRevE.60.7186} and, consequently, the overall evolution of materials toward equilibrium. In binary systems, particularly those involving solution and stoichiometric phases, the discrepancies in molar volumes introduce additional layers of complexity, influencing phase stability\cite{PhysRevE.72.011602}, nucleation kinetics\cite{Kumar2007, Warren2002}, and the formation of microstructures\cite{MOELANS2008268, WARREN1995689}.

The tailored phase field calculation becomes an invaluable approach in this context, offering a framework to explore how these molar volume differences manifest in the diffusion evolution of binary systems. This nuanced understanding is imperative for optimizing material properties, as it enables researchers to tailor their approaches based on the distinctive characteristics of each phase in systems. In metallurgy, for instance, predicting and controlling the microstructural evolution in alloys with varying molar volumes is pivotal for achieving desired mechanical properties. Meanwhile, manipulating molar volume differences in binary systems can influence phase stability, morphology, and electronic properties, opening avenues for designing advanced materials with tailored functionalities.

However, in most of the existing phase-field models, the common assumption is constant $V_m$ across diffusion zone\cite{Tripathi2021, Vach2012}. Some works also include molar volume difference by applying Pilling–Bedworth ratio to introduce molar volume difference between two distinct phases with constant molar volume.\cite{yang2011theoretical, ansari2021phase, zhao2022phase} However, real-world material systems are much more complicated to model. Hence, considering molar volume difference is necessary for more accurately predicting the microstructure evolution behaviors.

This article aims to delve into the core importance of molar volume differences in real-world material systems such as Al-Cu and Al-Mg, emphasizing their profound influence on the behavior of solution and stoichiometric phases. We will use four different molar volume settings in phase-field simulations: (1) uniform molar volume, (2) uniform molar volume inside the solution phase, (3) linear dependence on composition, and (4) non-linear dependence on composition. By doing so, we seek to provide a comprehensive computation foundation for researchers and practitioners to navigate the effect of molar volume differences of different phases, paving the way for advancements in material design and engineering across diverse applications.

\section{Model Description}
We introduce the molar volume difference based on the model from our previous work\cite{JI2022118007}, which focuses on modeling stoichiometric compounds. To describe the reaction to form a stoichiometric compound from the solution phase, an order parameter $\xi$ is defined as:
    
    \begin{equation}\label{eq1}
        \xi = 
        \begin{cases}
             0, & \quad \text{full solution phase without any reaction}\\
             1, & \quad \text{full precipitate phase with reaction completed}
        \end{cases}
    \end{equation}

The total free energy of the two-phase mixture can be formulated as:

    \begin{equation}\label{eq2}
        G = \int(g_{bulk}+g_{int}+g_{el})dV
    \end{equation}

where $g_{bulk}$, $g_{int}$, and $g_{el}$ are bulk, interfacial, and elastic strain free energy densities, respectively. Thus, the free energy can be written as a function of the order parameter $\xi$ and local composition $x_B$ in the solution.

\subsection{Bulk and Interfacial Energy Contribution}
We first redefine the local concentration $c$ as a function of $x_B$ and $\xi$, then formulate the bulk free energy $g_{bulk}$ as: 

    \begin{equation}\label{eq3}
        g_{bulk} = c(x_B,\xi)\mu^{tot}(x_B,\xi)
    \end{equation}

where $\mu^{tot}(x_B,\xi)$ is $\Bigl\{[1-h(\xi)][\mu_A(x_B)(1-x_B)+\mu_B(x_B)x_B]+h(\xi)\mu_{A_{v_A}B_{v_B}}^o\Bigr\}$  and $h(\xi)=6\xi^5-15\xi^4+10\xi^3$ is an interpolation function so that $h(0)=0$, $h(1)=1$ and $h^\prime(0)=h^\prime(1)=0$, and $\mu^\alpha=\mu_A (x_B )(1-x_B )+\mu_B (x_B ) x_B$ is the chemical potential, or molar Gibbs free energy of the matrix phase\cite{HU2007303, murray1985aluminium}. Note the local total composition $x_B$ in the system now becomes $x_B^{tot}=[1-h(\xi)] x_B+v_B h(\xi)$ where $v_B$ is the stoichiometric coefficient of B in the compound. With the definition of \(c(x_B,\xi)\) given by \(c(x_B,\xi) = c_A + c_B = \frac{1}{V_\alpha^m(1-h(\xi)) + V_{A_{v_A}B_{v_B}}^mh(\xi)}\), where \(V_\alpha^m\) is the molar volume of the solution phase, which may be a linear or nonlinear function of composition. Meanwhile, \(V_{A_{v_A}B_{v_B}}^m\) is the molar volume of the compound \(A_{v_A}B_{v_B}\), which is a constant. Thus, the total concentration  $c(x_B, \xi)$ is no longer a constant but a function of the mole fraction $x_B$ and the order parameter $\xi$. Meanwhile, the chemical potential of the compound, $\mu_{A_{v_A}B_{v_B}}^o$, remains constant under specific temperature and pressure conditions. Following the common practice of the phase-field model, we introduce a double-well potential energy function $g^{dw} (\xi)=\xi^2 (1-\xi)^2$, and a gradient energy contribution to introduce the contribution of interfacial energy: 

    \begin{equation}\label{eq4}
        g_{int} = wg^{dw}(\xi) + \frac{1}{2}\kappa(\nabla\xi)^2
    \end{equation}

where $w$ is the height of the double-well potential and $\kappa$ is the gradient coefficient. Since the contribution to the free energy above the common tangent in the two-phase region is automatically subtracted off, there is no need to introduce composition gradient terms in $g_{int}$.

\subsection{Elastic Strain Energy Contribution}
The formation of the stoichiometric compound in a solid solution also results in lattice mismatch between the solution and the compound which generates elastic strain energy. We define $\varepsilon_{ij}^0$ as the eigenstrain due to the compositional dependence of the lattice parameter of the solution phase and the lattice parameter difference between the stoichiometric precipitate and the solution matrix. Since the formulation of $\varepsilon_{ij}^0$ should be consistent with the composition- and phase-dependence of the molar volumes defined in previous sections, we write the dependence of eigenstrain as a function of composition and order parameter as 

    \begin{equation}\label{eq5}
        \varepsilon_{ij}^o = [1-h(\xi)]\varepsilon^{c}(x_B)\delta_{ij} + h(\xi)\varepsilon_{ij}^{00}
    \end{equation}

where $\varepsilon^c$ is the compositional strain due to the composition-dependent molar volume of the solution phase (originated from the size difference between A and B atoms), $\delta_{ij}$ is the Kronecker delta function, $\varepsilon_{ij}^{00}$ is the stress-free transformation strain (SFTS) obtained from the stress-free lattice parameters of the solution matrix phase and the stoichiometric compound phase, which accounts for the molar volume difference between the compound phase and the solution phase. The eigenstrain components should satisfy 

    \begin{subequations}\label{eq6ab}
        \begin{align}
            \det(I+\varepsilon^c) = \frac{V_\alpha^m(x_B)}{V_\alpha^m(x_B^0)}\\
            \det(I+\varepsilon^{00}) = \frac{V_{A_{v_A}B_{v_B}}^m}{V_\alpha^m(x_B^0)}
        \end{align}
    \end{subequations}

Where $I$ is the identity matrix, $\varepsilon^c$ is the compositional strain tensor with components $\varepsilon^c (x_B)\delta_{ij}$ and $\varepsilon^{00}$ is the SFTS tensor with components $\varepsilon_{ij}^{00}$, and $x_B^0$ is a reference composition in the solution phase. From Eq.(6a), we can immediately obtain

    \begin{equation}\label{eq7}
        \varepsilon^c(x_B) = \sqrt[3]{\frac{V_\alpha^m(x_B)}{V_\alpha^m(x_B^0)}}-1
    \end{equation}

The components of $\varepsilon^{00}$, however, depend highly on the orientation relationship and lattice correspondence between the matrix and the compound phases, which we will discuss in detail later for different materials systems.

By assuming linear elasticity, the local elastic strain energy density is given by

    \begin{equation}\label{eq8}
        g_{el} = \frac{1}{2}C_{ijkl}(\varepsilon_{ij}-\varepsilon_{ij}^o)(\varepsilon_{kl}-\varepsilon_{kl}^o)
    \end{equation}

where $C_{ijkl}$ is the elastic stiffness tensor, $\varepsilon_{ij}$ is the total strain obtained from the mechanical equilibrium condition $\nabla_j[C_{ijkl}(\varepsilon_{kl}-\varepsilon_{kl}^0 )]=0$. 

The total strain $\varepsilon_{ij}$  includes the contributions from homogeneous and heterogeneous deformations, i.e., $\varepsilon_{ij}=\varepsilon_{ij}+\delta\varepsilon_{ij}$ where the heterogeneous strain $\delta\varepsilon_{ij}$  is related to the local displacement $u$ via $\delta\varepsilon_{ij}=\frac{1}{2}(u_{i,j}+u_{j,i} )$. Following the micro-elasticity theory, the mechanical equilibrium equation $\nabla_j [C_{ijkl}(\varepsilon_{kl}-\varepsilon_{kl}^0 )]=0$ can be solved in Fourier space, from which we can obtain the local microstructure-dependent displacement $\mathbf{u}$,

    \begin{equation}\label{eq9}
        u_k(\mathbf{r}) = -\mathbb{i}\int\frac{d^3q}{(2\pi)^2}\frac{1}{q}\Omega_ik(\mathbf{n})\sigma_{ij
                        }^0(\mathbf{q})n_je^{\mathbb{i}q\cdot{r}}
    \end{equation}

where $\mathbb{i}=\sqrt{-1}$, $\mathbf{r}$ and $\mathbf{q}$ are the spatial coordinate vector in real space and Fourier space, respectively; $\mathbf{n=q/\mid{q}\mid}$ is a unit vector in Fourier space, $\sigma_{ij}^0 (\mathbf{q})=C_{ijkl} \widetilde\varepsilon_{kl}^{0}(\mathbf{q}), \widetilde\varepsilon_{kl}^{0}(\mathbf{q})$ is the total eigenstrain in Fourier space, and $\Omega_{ik}(\mathbf{n})=(C_{ijkl}n_jn_l)^{-1}$ is a Green function tensor.

The homogeneous strain $\bar\varepsilon_{ij}$ can be determined from the mechanical boundary conditions. In this study, we use constrained boundary conditions, i.e., $\bar\varepsilon_{ij}=0$. It should be noted that if the molar volumes of the phases are significantly different (e.g., $>$20\% difference), large deformation should be considered, which should include both the geometric nonlinearity and the plastic deformation. A few existing phase-field models\cite{zhao2022phase} have been developed for these considerations, although none of them have considered the stoichiometric formulation in our current work. Since the current work focuses on the thermodynamic description and bulk driving forces of systems involving stoichiometric phases and molar volume differences, these more rigorous considerations of the mechanical effects are out of our research scope, which we will explore in our forthcoming research.

\subsection{Governing Equations}
With the consideration of the bulk, interfacial, and elastic energy contributions, the governing phase-field evolution equations become

    \begin{subequations}
        \begin{align}
            \begin{split}\label{eq10a}
                \frac{\partial\xi}{{\partial}t} = -L_\xi\left(w\frac{{\partial}g^{dw}}{\partial\xi} - \kappa\nabla^2\xi + \frac{{\partial}g_{bulk}}{\partial\xi} + \frac{{\partial}g_{el}}{\partial\xi}\right)
            \end{split}\\
            \begin{split}\label{eq10b}
                \frac{{\partial}x_B}{{\partial}t} = \nabla\cdot\left[M\nabla\left(\frac{{\partial}g_{bulk}}{{\partial}x_B^{tot}}+ \frac{{\partial}g_{el}}{{\partial}x_B^{tot}}\right)\right]- \frac{{\partial}[h(\xi)(v_B-x_B)]}{{\partial}t}
            \end{split}
        \end{align}
    \end{subequations}

where $L_\xi$ is a reaction coefficient associated with stoichiometric reaction and M is the interdiffusion mobility.

With

    \begin{subequations}
        \begin{align}
            \begin{split}\label{eq11a}
                \frac{{\partial}g_{bulk}}{\partial\xi} &= c(x_B,\xi)\frac{{\partial}h}{\partial\xi}\\
                        &\quad\left\{\Delta\mu^r(x_B)-c(x_B,\xi)\cdot\mu^{tot}(x_B,\xi)\left[V_{A_{v_A}B_{v_B}}^m - V_\alpha^m(x_B) + \frac{\partial V_\alpha^m}{\partial x_B}(x_B-v_B) \right]\right\}
            \end{split}\\
            \begin{split}\label{eq11b}
                \frac{{\partial}g_{el}}{\partial\xi} &= -C_{ijkl}(\varepsilon_{ij}- \varepsilon_{ij}^0)\frac{\partial\varepsilon_{kl}^0}{\partial\xi}\\
                    &=-C_{ijkl}(\varepsilon_{ij}-\varepsilon_{ij}^0)\frac{{\partial}h}{\partial\xi}\left[\varepsilon_{kl}^{00}-\varepsilon^{c}(x_B)\delta_{kl}+\frac{\partial\varepsilon^c}{{\partial}x_B}(x_B-v_B)\delta_{kl}\right]
            \end{split}\\
            \begin{split}\label{eq11c}
                \frac{{\partial}g_{bulk}}{{\partial}x_b^{tot}} &= c(x_B,\xi)\left[\mu_B(x_B)-\mu_A(x_B)-c(x_B,\xi)\cdot\mu^{tot}(x_B,\xi)\cdot\frac{{\partial}V_\alpha^m}{{\partial}x_B} \right]
            \end{split}\\
            \begin{split}\label{eq11d}
                \frac{{\partial}g_{el}}{{\partial}x_b^{tot}} &= -C_{ijkl}(\varepsilon_{ij}-\varepsilon_{ij}^0)\frac{\partial\varepsilon_{kl}^0}{{\partial}x_B^{tot}}
                 =-C_{ijkl}(\varepsilon_{ij}-\varepsilon_{ij}^0)\frac{\partial\varepsilon^c}{{\partial}x_B}\delta_{kl}
            \end{split}
        \end{align}
    \end{subequations}

    where $\Delta\mu^r(x_B) = \mu^0_{A_{v_A}B_{v_B}} - v_A\mu_A - v_B\mu_B $ is the chemical potential change for the stoichiometric reaction.

It should be noted that the chemical potentials $\mu_B(x_B)$ and $\mu_A(x_B)$ typically contain logarithm functions of composition, which causes numerical issues when the local composition is lower than 0 or higher than 1. To alleviate this issue, as has been discussed in our previous work\cite{JI2022118007}, we can evolve $Y= \ln\frac{x_B}{1-x_B}$ in Eq.(10b) rather than directly evolve $x_B$. The evolution equation now becomes

    \begin{equation}\label{eq12}
        \frac{e^Y}{(1+e^Y)^2}\frac{{\partial}Y}{{\partial}t} = \nabla\cdot\left[M\nabla\left(\frac{{\partial}g_{bulk}}{{\partial}x_B^{tot}}+\frac{{\partial}g_{el}}{{\partial}x_B^{tot}}\right)\right]-\frac{{\partial}[h(\xi)(v_B-x_B)]}{{\partial}t}
    \end{equation}
The intermediate parameter $Y$ can be converted to composition via $x_B=\frac{e^Y}{1+e^Y}$.    

\section{Model Applications and Discussions}
\subsection{Model Setup}
\begin{figure}
        \centering
        \includegraphics[width=\textwidth]{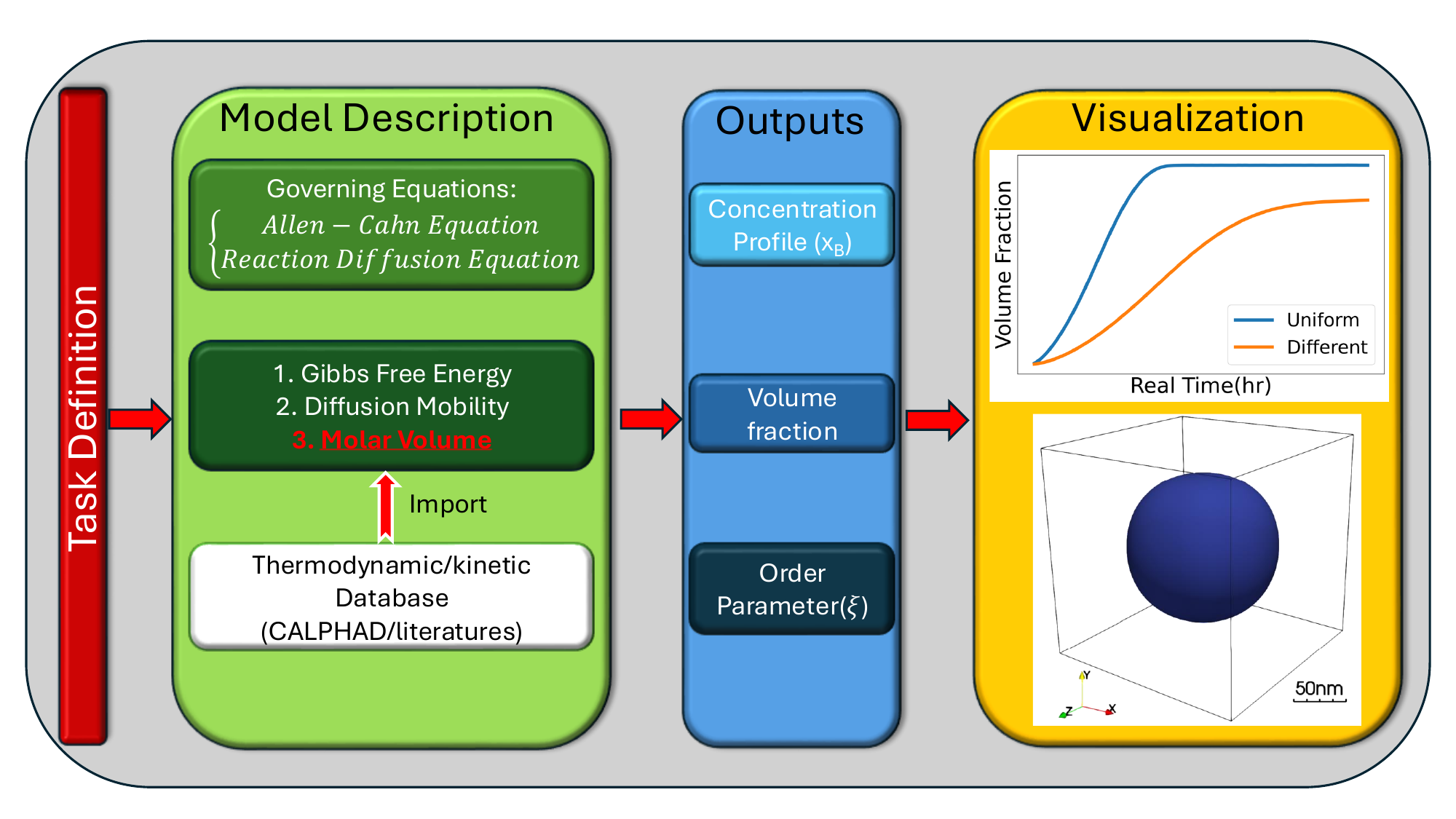}
        \caption{Illustration of work flow}
        \label{flow_schematic}
    \end{figure}

In this section, we perform 3-D simulations of the growth of stoichiometric phases with different molar volume settings and their consistent elastic strain energies. Model performances are evaluated by comparing the volume fractions of precipitation with the theoretical values obtained by the lever rule. The lever rule provides an upper limit of precipitation  fraction. Fig. \ref{flow_schematic} demonstrates the workflow for model applications from acquiring key parameter values from the database(CALPHAD or literature) into governing equations then visualizing morphology and performing precipitate volume fraction analysis with theoretical values.

\subsection{Application to $\theta^\prime$ Precipitation in Al-Cu Alloy}
We first consider the simulations of the precipitation of stoichiometric $\theta^\prime$-Al$_2$Cu (with stoichiometric coefficients of $v_{Al}$=2/3 and $v_{Cu}$=1/3) from the $\alpha$ solid solution in Al-Cu alloys assuming (1) uniform molar volume  $V_\alpha^m$= $V_{\theta^\prime}^m$ = 9.920$\times10^{-6}$ $m^3/mol$; (2) uniform molar volume only inside the solution phase with $V_\alpha^m$=9.920$\times10^{-6}$ $m^3/mol$ and $V_{\theta^\prime}^m$=9.437$\times10^{-6}$ $m^3/mol$; (3) molar volume of solution phase has linear dependence on composition as $V^m_{\alpha} = (1-x_{Cu})V^0_{Al}+x_{Cu}V^0_{Cu}$ where $V_{Al}^0$=9.920$\times10^{-6}$ $m^3/mol$($\alpha_{Al}=4.04 \mathring{A}$\cite{davey1925precision}), $V_{Cu}^0$=6.892$\times10^{-6}$ $m^3/mol$($\alpha_{Cu}=3.57 \mathring{A}$\cite{wei1987first}) and $V_{\theta^\prime}^m$=9.437$\times10^{-6}$ $m^3/mol$ from Materials Project\cite{jain2013commentary}; and (4) a comprehensive form of molar volume of solution phase has an additional nonlinear function of composition multiplied after the linear term in case(3) with the expression as Eq. \ref{eq13} from molar volume database of Al-Cu.\cite{HUANG2020101693} 

\begin{equation}\label{eq13}
    V^m_\alpha = [(1-x_{Cu})V^0_{Al}+x_{Cu}V^0_{Cu}+x_{Cu}(1.-x_{Cu})V^0_0]\exp[(1-x_{Cu})V^{Cu}_{Al}+x_{Cu}V^{Cu}_{Cu}+x_{Cu}(1-x_{Cu})V^A_0]
\end{equation}

where $V^{Cu}_{Al}$ and $V^{Cu}_{Cu}$ are integrated thermal expansion, $V^A_0$ and $V^0_0$ are first-order interaction parameters.

A cubic system with 160 grids and a grid spacing of 1nm is employed. The initial matrix composition $x^0_{Cu}$ is set at 1.5at.\%. Isothermal aging is performed at 463\, K with the reduced time step \( \Delta t^* \) of 0.02, and Cu impurity diffusivity $D_{Cu}$ in FCC Al is $8.88 \times 10^{-5} \times \exp\left(\frac{-133900}{RT}\right) \, \text{m}^2/\text{s}$ obtained by Arrhenius relation.\cite{du2003diffusion} Therefore, a corresponding real-time step of $\Delta t = 2.88s$ can be acquired. Both expressions of chemical potentials of each phase and molar volumes are taken from the previous work\cite{HU2007303, HUANG2020101693}. The interfacial energies, elastic constants, and SFTS are taken from our previous work\cite{JI201884}.

\subsubsection{Constant molar volumes: uniform and different}
\begin{figure}
    \centering
    \includegraphics[width=\textwidth]{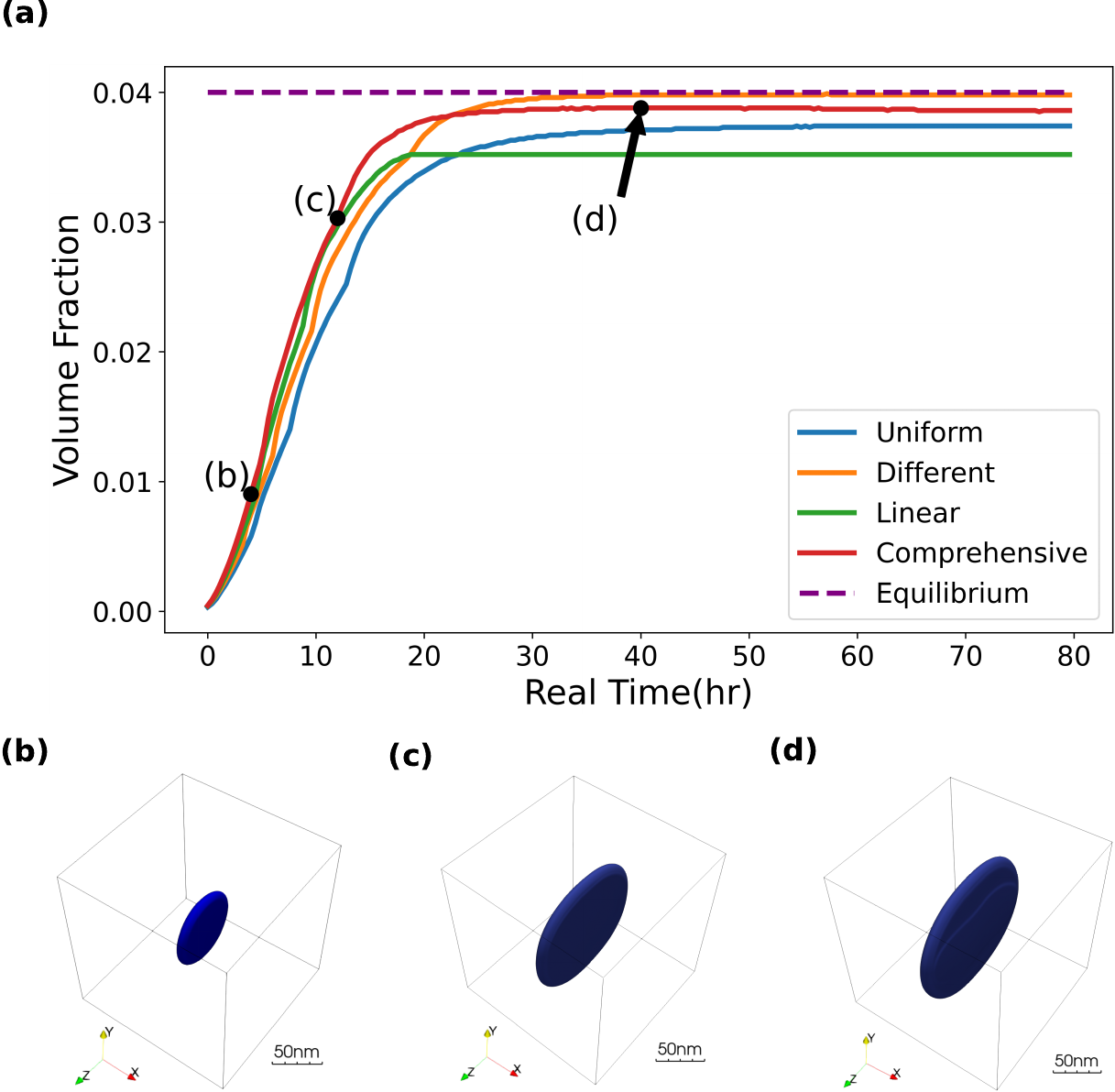}
    \caption{\textbf{(a)}volume fraction of $\theta^\prime$-Al$_2$Cu for each molar volume setting in 3-dimension. \textbf{(b)$\sim$(d)}morphology of Al$_2$Cu during stimulation at 4, 12, and 40 (hr) respectively}
    \label{fig:3DAlCu}
\end{figure}

Fig. \ref{fig:3DAlCu}(a) shows the full $\theta^\prime$-Al$_2$Cu precipitation history with the final volume fraction of 3.74\%, 3.99\%, 3.52\%, 3.88\%, and 4.0\%\cite{HU2007303} for each molar volume setup of uniform, different, solution phase linear dependence on composition, the comprehensive setup with nonlinear composition dependence, and the equilibrium value obtained by lever rule respectively. The growth rates in both the uniform and different cases are lower than those in the linear and comprehensive setups due to the constant molar volume in the solution phase, which leads to a smaller bulk driving force for composition evolution as $\frac{{\partial}V_\alpha^m}{{\partial}x_B}$=0 in Eq. \ref{eq11c}. Similar scenarios apply to the bulk driving force for order parameter evolution in Eq. \ref{eq11a}. For uniform cases, since $V_{A_{v_A}B_{v_B}}^m - V_\alpha^m(x_B) + \frac{\partial V_\alpha^m}{\partial x_B}(x_B-v_B) = 0$, the bulk driving force for order parameter evolution simply depends on the chemical potential of forming the $\theta^\prime$-AlCu compound only. Different molar volume setups also lower the bulk driving force for order parameter evolution, but $V_{A_{v_A}B_{v_B}}^m - V_\alpha^m(x_B) \ne 0$. Therefore, both uniform and different setups decrease the total bulk driving force in either composition or order parameter evolution, causing a slower precipitation rate, as shown in Fig. \ref{fig:3DAlCu}(a).

Since the lever rule does not assume the same molar volume between two phases, the uniform set will have a larger deviation from equilibrium than the different molar volume setup. In addition to the lever rule, previously discussed bulk driving force contributions in both cases also contribute to the final precipitate volume fraction. In the uniform setting, the only contributor is the chemical potential of the $\theta^\prime$-Al$_2$Cu compound, while the different molar volume setting has an additional fluctuation term contributed by the $V_{A_{v_A}B_{v_B}}^m - V_\alpha^m(x_B)$ term in Eq. \ref{eq11a}. 


Unlike the lever rule, which solely depends on bulk free energy, the approach in this study considers both bulk free energy and elastic strain energy. The elastic strain energy from composition difference contributes to uniform and different final precipitate volume fractions. 



\subsubsection{Functional molar volumes: linear and comprehensive}
The scenario for functional molar volume settings differs from the constant assumption due to the composition-dependent molar volume in the matrix phase. As shown in Fig. \ref{fig:3DAlCu}(a), both linear and comprehensive settings exhibit a more rapid growth rate than the constant molar volume. This is an outcome of the larger bulk free energy driving force contributed by  $V_{A_{v_A}B_{v_B}}^m - V_\alpha^m(x_B) + \frac{\partial V_\alpha^m}{\partial x_B}(x_B-v_B) \neq 0$ in Eq. \ref{eq11a}.

In the linear setting, $\frac{\partial V_\alpha^m}{\partial x_B}$ is a constant where $V_{Cu}^m - V_{Al}^m = -3.028 \times 10^{-6} \text{m}^3/\text{mol}$. This makes the driving force of the precipitate phase more negative, thereby leading to a larger driving force and faster growth rate. In contrast, for the comprehensive setting, the gradient term $\frac{\partial V_\alpha^m}{\partial x_B}$ is a function of composition that falls within the range of $-1.362 \times 10^{-6} \text{m}^3/\text{mol}$ to $-4.379 \times 10^{-6} \text{m}^3/\text{mol}$ across the entire composition range. Similar to the linear setting, the composition-dependent matrix phase molar volume provides a larger bulk free energy driving force, resulting in rapid precipitate growth, as shown in Fig. \ref{fig:3DAlCu}(a).

As mentioned in the previous section, the elastic strain energy decreases the overall precipitate volume fraction to a level lower than the equilibrium value obtained from the lever rule. According to Eq. \ref{eq8}, the elastic strain under the linear molar volume assumption is $-1.550 \times 10^{-5}$, which decreases the elastic strain driving force in Eq. \ref{eq11b}. The comprehensive form of molar volume results in an elastic strain of $-2.374 \times 10^{-5}$, which has an identical effect on the driving force as the linear form. However, since the magnitude of the elastic strain is so small, its contribution to the final precipitate volume fraction may be negligible.


Fig. \ref{fig:3DAlCu}(b)$\sim$(d) shows the morphology of $\theta^\prime$-Al$_2$Cu resulting from a comprehensive molar volume setting at real-time steps of 4, 12, and 40 hours, respectively. The plate-like $\theta^\prime$-Al$_2$Cu precipitated from the calculation has an identical structure to the previous experimental results\cite{biswas2011precipitates}, which was contributed from anisotropic interfacial energy combined with elastic strain energy and bulk driving force.

\subsection{Application to $\beta$-precipitates in Al-Mg Alloys}
We consider the stoichiometric compound of $\beta$-\ce{Al_{140}Mg_{89}} in the Al-Mg alloy system with stoichiometric coefficients of $v_{Al}=140/229$ and $v_{Mg}=89/229$. The molar volume of the solution matrix phase was taken from the Al-Mg molar volume database. The molar volume of $\beta$-\ce{Al_{140}Mg_{89}} is $V_{\beta}^m$=1.13×$10^{-5}$ $m^3/mol$\cite{zhong2005contribution}. Two assumptions were applied on 3-D simulation with (1) uniform molar volume:$V_\alpha^m$ = $V_{\beta}^m$=1.13×$10^{-5}$ $m^3/mol$; (2) different molar volume: $V_\alpha^m$=9.92×$10^{-6}$ $m^3/mol$ and $V_{\beta}^m$=1.13×$10^{-5}$ $m^3/mol$.

The same 3-D cubic simulation system was used in $\beta$ -\ce{Al_{140}Mg_{89}} as in $\theta^\prime$ -Al$_2$Cu. The initial Mg composition $x^0_{Mg}$ in the matrix phase was 10.0at.\% and the isothermal aging temperature was set at 300K. The Mg impurity diffusivity in FCC Al along grain boundaries is $2.24 \times 10^{-1} \times \exp\left(\frac{-142797}{RT}\right) \, \text{m}^2/\text{s}$.\cite{koju2020atomistic}

\begin{figure}
    \centering
    \includegraphics[width=\textwidth]{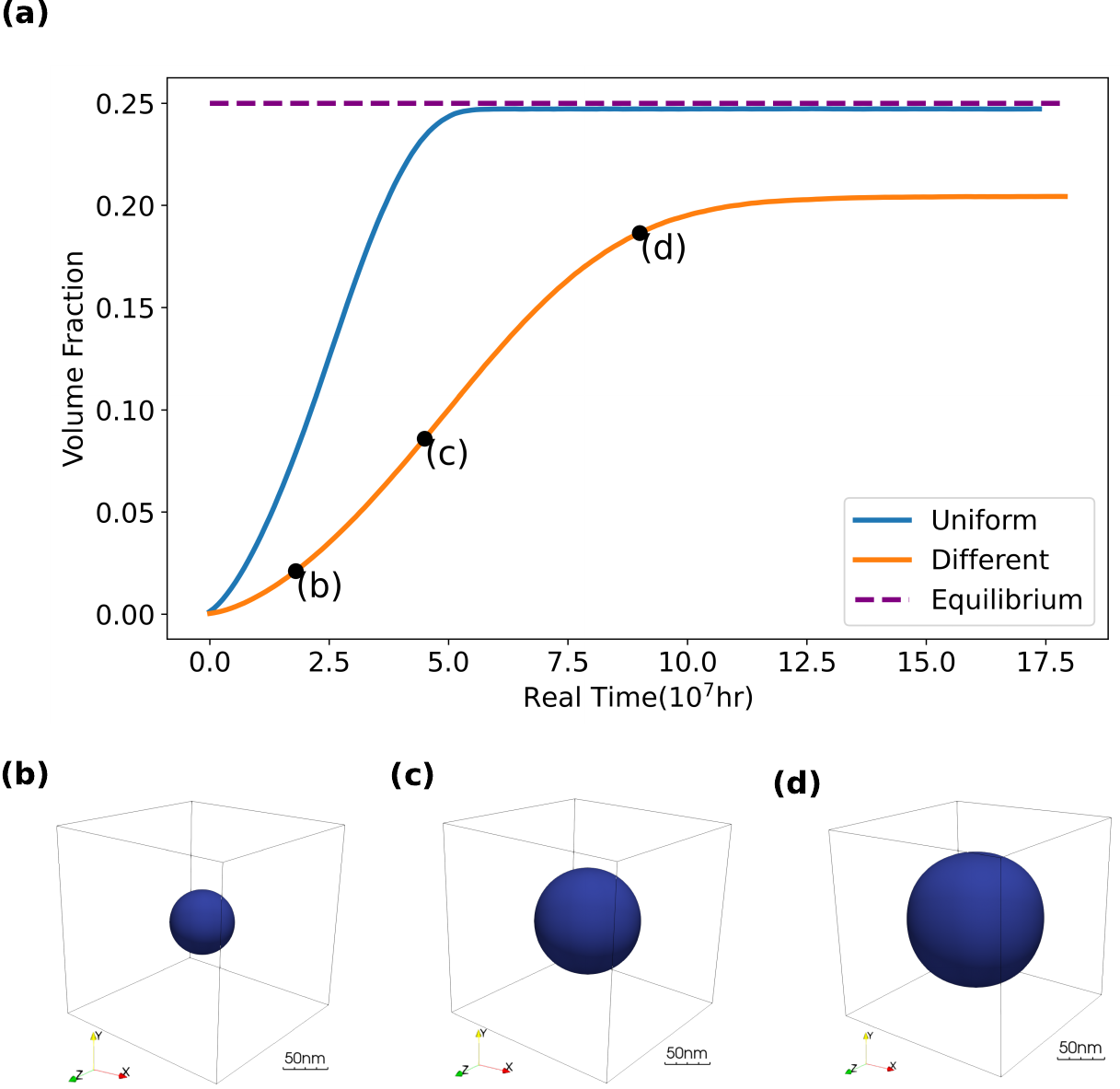}
    \caption{\textbf{(a)}volume fraction of $\beta$-AlMg for each molar volume setting in 3-dimension. \textbf{(b)$\sim$(d)}morphology of $\beta$-AlMg during simulation at 2, 4, and 8($\times10^7$hr) respectively}
    \label{fig:3DAlMg}
\end{figure}

The final precipitation volume fraction is 24.7\%, 20.4\%, and 25.0\% for uniform and different molar volume assumptions and equilibrium, respectively. The final precipitate volume fraction in the different molar volume setup shows a larger deviation from the theoretical equilibrium than the uniform setup. Since the molar volume of $\beta$-\ce{Al_{140}Mg_{89}} in the different molar volume setup is much larger than that of the solution matrix phase, which results in $V_{A_{v_A}B_{v_B}}^m - V_\alpha^m(x_B) > 0$, the bulk driving force is less negative than in the uniform case, according to Eq. \ref{eq11a}. Therefore, the overall bulk driving force in the uniform setup is larger than in the different molar volume setup, and the precipitation rate is lower in the different molar volume setup, as shown in Fig. \ref{fig:3DAlMg}(a). Moreover, since the molar volume of the precipitate phase is much larger (approximately 13\%) than that of the solution matrix phase, a significant elastic strain will hinder precipitation and further lower the equilibrium precipitate volume fraction. On the other hand, the uniform molar volume setup only has a compositional strain contribution, making the final equilibrium volume fraction closer to the theoretical equilibrium.

Fig. \ref{fig:3DAlMg}(b)–(d) illustrates the precipitate morphology of $\beta$-\ce{Al_{140}Mg_{89}} under isothermal conditions with different molar volume assumptions and at three different calculation times. The spherical shape of the precipitate was obtained and maintained either at the initial stage or near final equilibrium. This precipitate morphology aligns with experimental characterization results, which show a similar Mg composition at 8.9 at\%.\cite{zhou2021making} Despite the isothermal temperature from the experimental discovery being significantly higher than our calculation setup, the precipitate morphologies are identical, which validates our molar volume assumption using the phase field approach.

\section{Conclusion}
In this study, various molar volume hypotheses have been investigated concerning their influence on precipitate kinetics from both quantitative and qualitative aspects. The $\theta^\prime$-Al${_2}$Cu precipitate shows much smaller deviation from theoretical equilibrium under different and comprehensive molar volume assumptions. The $\beta$-\ce{Al_{140}Mg_{89}} precipitate is better aligned with the theoretical volume fraction under the uniform assumption than in the varied case, which is a result of the large molar volume gap between the solution matrix phase and the precipitate phase. Both the Al-Cu and Al-Mg systems have bulk driving force and compositional strain contributing to the precipitate kinetics due to the various molar volume hypotheses. This study also validated the precipitate morphology with experimental evidence to further examine the feasibility of our approach. 

These results lay a solid foundation for the microstructure evaluation of the stoichiometric compound formation accompanied with molar volume change using phase-field approaches, which provide robust guidance for future experimental investigations on different alloy systems.

\section{Acknowledgement}
The authors acknowledge the start-up fund from The Ohio State University. Computations were supported by the Ohio Supercomputer Center.

\bibliographystyle{elsarticle-num-names} 
\bibliography{main} 
\end{document}